\documentclass[sigconf,nonacm]{acmart}
\AtBeginDocument{%
  \providecommand\BibTeX{{%
    \normalfont B\kern-0.5em{\scshape i\kern-0.25em b}\kern-0.8em\TeX}}}

\usepackage{listings}
\usepackage{subfigure}
\usepackage{amsmath}
\usepackage{adjustbox}
\usepackage{graphicx}
\usepackage[linesnumbered,boxed]{algorithm2e}
\usepackage{caption}
\usepackage{balance}

\SetKwInput{KwInput}{Input}
\SetKwInput{KwOutput}{Output}

\setcopyright{none}
\begin{document}

\title{MKPipe: A Compiler Framework for Optimizing Multi-Kernel Workloads in OpenCL for FPGA}

\settopmatter{printacmref=false}


\author{Ji Liu}
\email{jliu45@ncsu.edu}
\orcid{0000-0002-5509-5065}
\affiliation{%
  \institution{North Carolina State University}
  \city{Raleigh}
  \state{North Carolina}
  \postcode{27606}
}

\author{Abdullah-Al Kafi}
\email{akafi2@ncsu.edu}
\affiliation{%
  \institution{North Carolina State University}
  \city{Raleigh}
  \state{North Carolina}
  \postcode{27606}
}

\author{Xipeng Shen}
\email{xshen5@ncsu.edu}
\affiliation{%
  \institution{North Carolina State University}
  \city{Raleigh}
  \state{North Carolina}
  \postcode{27606}
}

\author{Huiyang Zhou}
\email{hzhou@ncsu.edu}
\orcid{0000-0003-2133-0722}
\affiliation{%
  \institution{North Carolina State University}
  \city{Raleigh}
  \state{North Carolina}
  \postcode{27606}
}

\begin{abstract}
OpenCL for FPGA enables developers to design FPGAs using a programming model similar for processors. Recent works have shown that code optimization at the OpenCL level is important to achieve high computational efficiency. However,  existing works either focus primarily on optimizing single kernels or solely depend on channels to design multi-kernel pipelines. In this paper, we propose a source-to-source compiler framework, MKPipe, for optimizing multi-kernel workloads in OpenCL for FPGA. Besides channels, we propose new schemes to enable multi-kernel pipelines. Our optimizing compiler employs a systematic approach to explore the tradeoffs of these optimizations methods. To enable more efficient overlapping between kernel execution, we also propose a novel workitem/workgroup-id remapping technique. Furthermore, we propose new algorithms for throughput balancing and resource balancing to tune the optimizations upon individual kernels in the multi-kernel workloads. Our results show that our compiler-optimized multi-kernels achieve up to 3.6x (1.4x on average) speedup over the baseline, in which the kernels have already been optimized individually.

\end{abstract}






\maketitle

\section{Introduction}
FPGAs are reprogrammable devices that can be configured to perform arbitrary logic operations. Given their high energy efficiency, FPGAs have become an attractive accelerator platform for high performance computing \cite{zheng2014design,zhang2015optimizing}. Traditional FPGA design is through register-transfer level (RTL) Hardware Description Languages (HDL) such as Verilog and VHDL, which is time-consuming and unfriendly to software programmers. High-level synthesis (HLS), especially OpenCL for FPGAs \cite{intel,Xilinx}, offers a high-level abstraction to enable software developers to program FPGAs similar to processors and makes it possible to port the existing OpenCL code developed for CPUs or GPUs to FPGAs.

Similar to processor-based computing platforms, in order to achieve high performance, it is important to optimize the OpenCL code for FPGAs \cite{bestpracticeguide}. Prior works \cite{zohouri2016evaluating,gautier2016spector,jia2016tuning,zhang2017improving} have shown that optimized OpenCL code utilizes the FPGA device more effectively and results in competitive designs compared to HDL-based designs \cite{hill2015comparative}. However, existing works on optimizing OpenCL code for FPGAs mainly focuses on single kernels. Zohouri et al. \cite{zohouri2016evaluating} evaluated and optimized the OpenCL kernels in the Rodinia benchmark suite, but they only proposed single-kernel optimizations and did not consider concurrent execution among multiple kernels. Gautier et al. \cite{gautier2016spector} presented an OpenCL FPGA benchmark suite and similar single-kernel optimization approaches had been employed. 

The Intel OpenCL for FPGA programming and optimization guide \cite{programmingguide} introduces channels/pipes as the key mechanism for passing data between kernels and enabling pipelining/concurrent execution across the kernels. There are a few works leveraging channels to stream data across multiple kernels for specific applications\cite{wang2015study,zohouri2018combined}. However, channels have a strict limitation on the producer and consumer. As stated in the programming guide, "A kernel can read from the same channel multiple times. However, multiple kernels cannot read from the same channel." 
As a result, it is difficult to use channels for kernels with complex producer-consumer relationship. 
In this paper, we propose a novel compiler framework for optimizing multi-kernel workloads in OpenCL for FPGA. 

In our development of this optimizing compiler, we first study the multi-kernel workloads from the existing OpenCL for FPGA benchmark suites, including Rodinia\cite{zohouri2016evaluating}, Spector\cite{gautier2016spector} and OpenDwarf\cite{verma2016opendwarf}. We find that coding in multi-kernels has several advantages over a single monolithic one, including design modularity, code reuse, and optimization flexibility. Our experiments, however, highlights that these multi-kernel applications suffer from low FPGA resource utilization. The fundamental reason is that although individual kernels have been tuned with optimizations such as loop unrolling, SIMD, compute unit replication, etc., and multiple kernels are synthesized to co-reside on the same FPGA simultaneously, the kernels are executed one after another in a sequential manner due to data dependencies across the kernels. Therefore, there is no concurrent kernel execution (CKE) and only part of the FPGA is active at any time, resulting in low effective resource utilization. 

In our paper, we propose a compiler scheme to optimize different types of multi-kernel workloads. The compiler takes the host code, the naive kernel code, and the profiling data of the naive kernels as input and outputs the optimized kernel code and associated host code. The naive kernel code means that it does not have any device-specific or platform-specific optimizations. The compiler first derives the kernel data flow graph from the host code. Then the compiler analyzes the producer-consumer relationship among kernels. Based on the type of the producer-consumer relationship and the profiling information of the naive kernels, the compiler classifies a workload into different categories and performs optimizations to enable multi-kernel pipelines accordingly. 
In the next step, the compiler fine-tunes optimizations for individual kernels to balance the throughput and/or the resource consumption among the kernels. Then, the compiler explores the option of splitting the multi-kernels into separate FPGA bitstreams, which trades off the re-programming overhead for improved performance of individual kernels. 

We conduct our experiments using a Terasic's DE5-Net board with Altera OpenCL SDK18.1. The experimental results show that our optimizing compiler can effectively improve performance. The optimized multi-kernels achieve up to 3.6x (1.4x on average) speedup over those in the benchmark suites, in which each kernel has been optimized individually.

In summary, our contributions in this work include:

\begin{itemize}
\item We propose a compiler framework for optimizing multi-kernel workloads in OpenCL for FPGA. To our knowledge, this is the first optimizing compiler for multi-kernels in OpenCL for FPGA. 
\item
We analyze the trade offs among different CKE approaches and propose a novel systematic compiler optimization scheme to enable multi-kernel pipelines.
\item We propose novel algorithms to balance the throughput and/or resource consumption among the kernels in a multi-kernel workload. Such a kernel balancing process has not been discussed in previous works in OpenCL for FPGA. 
\item We devise a scheme to explore bitstream splitting, which separates multiple kernels into more than one bitstream so as to enable more aggressive optimizations for individual kernels. 
\end{itemize}


\section{BACKGROUND}
Open Computing Language (OpenCL) is an open standard for parallel computing across heterogeneous platforms\cite{khronous}. The key to the OpenCL programming model is data-level parallelism (DLP). In a user-defined kernel function, each workitem performs operations on different data items based on its identifier (id). Multiple workitems in the same workgroup can communicate through local memory.

Intel FPGA SDK for OpenCL is designed for executing OpenCL kernels on FPGAs. It supports two different kernel modes: single workitem and NDRange. In the single workitem mode, the OpenCL compiler leverages parallel loops in the kernel code and converts loop-level parallelism into pipeline-level parallelism (PLP) by synthesizing the hardware from the loop body and pipelining independent loop iterations. In the NDRange mode, DLP is converted to PLP by synthesizing the kernel code into a pipeline (aka compute unit) and pipelining independent workitems. To improve the throughput of the pipeline resulting from the kernel code, typical single-kernel optimization techniques include loop unrolling and vectorization/SIMD to deepen and widen the pipeline, as well as compute unit replication to duplicate the pipelines \cite{programmingguide}. For single workitem kernels, shift registers are a commonly used pattern to improve hardware efficiency. It has been shown that some kernels achieve better performance with the single workitem mode while others prefer the NDRange mode \cite{jia2016tuning}. 

The OpenCL for FPGA model has four types of memory. Global memory resides in the off-chip DDR memory, which has high latency and the bandwidth is shared among all the compute units. Local memory is implemented using on-chip registers or block RAMs depending on the size and has low latency and high bandwidth. Constant memory also resides in the DDR memory but the constant data can be loaded into an on-chip cache that is shared by all work-items. Private memory is usually implemented as registers and has very low access latency. Similar to GPUs, software-managed local memory is commonly used with loop tiling/blocking to overcome the memory bottleneck while the hardware managed caches are synthesized to leverage the runtime data locality.

According to the programming and optimization guide of OpenCL for FPGA \cite{programmingguide}, channels/pipes are the main mechanism for passing data between kernels and enabling pipelining or concurrent execution across the kernels. Channels are on-chip FIFO buffers and there are two types of channels: blocking and non-blocking. For blocking channels, a read/write operation stalls when the channel is empty/full. For a non-blocking channel, a read/write always proceeds but has a return flag indicating whether the operation succeeds. As channels are the only approach discussed in the programming and optimization guide for concurrent kernel execution (CKE), it is no surprise that the prior works \cite{wang2015study,wang2017pipecnn,zohouri2018combined}, solely depend on channels to optimize their target multi-kernel applications.

\section{RELATED WORK}
The optimization techniques discussed in recent works on OpenCL for FPGA are mainly single-kernel optimizations. Besides the work by Zohouri et al.\cite{zohouri2016evaluating} and Gautier et al. \cite{gautier2016spector}, a more recent work by Zohouri et al. \cite{zohouri2018combined} proposed additional single-kernel optimizations, including loop collapsing to reduce resource consumption and exit condition optimization to reduce the logic critical path. 

Several previous works have discussed multi-kernel designs using channels for data streaming. Wang et al. \cite{wang2015study} studied the effect of using channels on a data partitioning workload. 
Yang et al.\cite{yang2017opencl} employed channels to implement a molecular dynamics application. 
Wang et al.\cite{wang2017pipecnn} designed an FPGA accelerator for convolution neural networks, which consists of a group of OpenCL kernels connected with channels. Although these prior works leverage multi-kernel pipelines, none of them goes beyond channels.

Sanaullah et al.\cite{sanaullah2018empirically} proposed an empirically guided optimization framework in OpenCL for FPGA and the goal is to best utilize the OpenCL compiler. In one step of the optimization, task-level parallelism from the single-kernel code is converted to multiple kernels connected with channels. In the next step, multiple kernels are converted back into a single kernel. Such exploration gives the compiler more room to generate different optimized code. And their observation was that channels often result in poor performance, which is kind of expected as their target workloads are single kernels.

In a recent work, Shata et al. \cite{shata2019optimized} studied the use of local atomic operations and other optimization methods. They also discussed the effectiveness of compiling multiple kernels into multiple bitstreams. However, they dismissed this option given the high reprogramming overhead and recommended to integrate the kernels in the same bitstream file.

Multi-kernel pipelines have also been investigated for GPUs\cite{zheng2017versapipe,steinberger2014whippletree}. Although both FPGA and GPUs use OpenCL, they are fundamentally different in implementing multi-kernel pipelines. We detail the differences in two multi-kernel pipeline models in Section~\ref{sec:comparison with GPU model}.

\section{Motivation}
\subsection{Concurrent Kernel Execution (CKE) on FPGA}
\label{sec:CKE on FPGA}
We first study the multi-kernel workloads from the existing OpenCL for FPGA benchmark suites, including Rodinia\cite{zohouri2016evaluating}, Spector\cite{gautier2016spector} and OpenDwarf\cite{verma2016opendwarf}. These multi-kernel workloads in our study share a common implementation that all the different kernels in the same workload are synthesized into a single bitstream, thereby co-residing on the same FPGA chip. The main reason is to eliminate the FPGA re-programming overhead. If there is data dependence among the kernels, the kernel invocations are sent to the same command queue, which imposes global synchronization among the invocations. The advantage of this approach is that it ensures correctness easily. The disadvantage is that the sequential execution of multiple kernels may lead to poor resource utilization since only the hardware corresponding to one kernel is active at a time.

\begin{figure}[htbp]
  \centering
  \includegraphics[height=2cm,width=8cm]{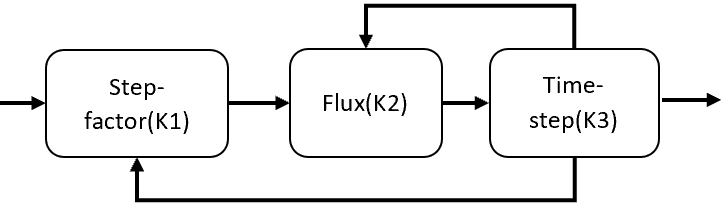}
  \caption{Kernel data flow graph of the CFD benchmark}
  \label{fig:cfd_invocation}
\end{figure}

To quantify the FPGA resource utilization, we propose a metric, effective resource utilization (ERU), for each kernel. As shown in Eq. \ref{effectiveresource}, it is defined as the maximum usage among different types of resources of the FPGA chip, including both the static resources: adaptive loop-up tables (ALUTs), dedicated logic registers (FFs), DSP blocks, RAMs blocks, and the dynamic resource: DRAM bandwidth. The static resource usage is computed as the percentage of the resource consumed by the kernel. The DRAM bandwidth usage is the ratio of the utilized bandwidth over the peak one when the kernel is active. The reason for the maximum is to capture the effect of the critical resource. This way, low effective resource utilization indicates that there is room available in the hardware for performance improvement.
\begin{equation}
\label{effectiveresource}
ERU = Max(U_{ALUT}, U_{FF}, U_{RAM}, U_{DSP},U_{BW})
\end{equation}
Here we use a case study on the benchmark CFD to illustrate the low ERU problem. CFD contains three kernels with the kernel data flow graph shown in Figure~\ref{fig:cfd_invocation}. Each kernel needs data from the previous one. 

\begin{figure}[htbp]
  \centering
  \includegraphics[height=6.5cm,width=8.2cm]{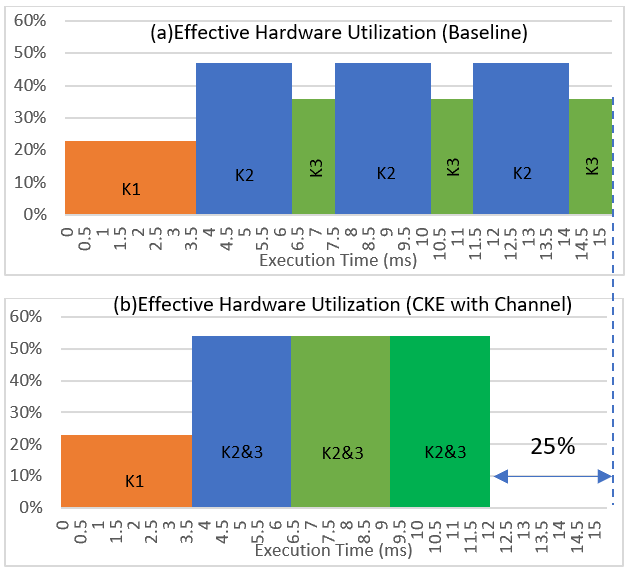}
  \caption{Effective resource utilization of CFD}
  \label{fig:cfdERU}
\end{figure}

 Figure~\ref{fig:cfdERU}a shows the ERU over time of the CFD benchmark. As the kernels are executed sequentially, we can visualize ERU as a stepwise function based on the order of kernel invocation and the execution time of each kernel. Although these kernels have been optimized individually, the overall ERU is low. In comparison, when we enable concurrent execution between K2 and K3, 
 the execution time is reduced as shown in Figure~\ref{fig:cfdERU}b. Furthermore, CKE using kernel fusion or channels may free up some hardware resource. For example, as shown in Figure~\ref{fig:cfdERU}b, the RAM usage of K2\&3 is less than the aggregated usage of K2 and K3. The reasons will be discussed in Section~\ref{sec:buidlingCKE}. Such freed up resource enables more opportunities for single-kernel optimizations. 
\subsection{Comparison with pipeline execution models for GPU}
\label{sec:comparison with GPU model}

The GPU multi-kernel pipeline programming frameworks \cite{steinberger2014whippletree,zheng2017versapipe} have been developed recently. These frameworks leverage multiple Streaming Multiprocessors (SMs) on a GPU and may schedule different pipeline stages on different SMs. This is possible as each SM is a programmable processing unit and can execute different kernels. In contrast, an FPGA has fixed functionality once it is synthesized. 
Therefore, the multi-kernel programming frameworks for GPU are not directly applicable for FPGAs. Here we dissect the five GPU pipeline execution models used in VersaPipe \cite{zheng2017versapipe} and analyze their similarities and differences compared to pipeline execution models for FPGA.

The first GPU pipeline execution model is "Run to completion (RTC)". This execution model combines all stages of a pipeline into a single kernel, which is similar to the kernel fusion method discussed in Section~\ref{sec:fusion}. The limitation of this model is that it does not support global synchronization between stages.

The second GPU execution model is "Kernel by kernel (KBK)". In this model, multiple kernels are used and the kernels are executed one after the other. This model is the same as the baseline in the benchmark suites we studied. The limitation is that there is no concurrent kernel execution as discussed in Section~\ref{sec:CKE on FPGA}.

The third GPU execution model is "Megakernel". Megakernel organizes pipeline computations into a huge kernel and each stage is scheduled with a software scheduler. The persistent thread technique~\cite{aila2009understandingPersistent,gupta2012persistent} is used to implement Megakernel. These persistent threads fetch data from a shared queue and run the corresponding pipeline stage upon the data. After each stage, the produced data are sent back to the same queue for subsequent processing.
We tried to implement the Megakernel design on FPGA. However, the OpenCL for FPGA compiler is unable to handle this type of kernel and cannot construct hardware based on the OpenCL code. The main problem is the switch statement which chooses among different pipeline stages. The compiler regards these kernels as "FPGA-unfriendly". Furthermore, there are additional drawbacks for this model. First, the scheduler requires extra hardware resources. Second, the data communication between each stage is based on a shared queue. Although we can implement this shared queue using local memory on FPGA, this queue becomes a bottle neck due to its high number of read and write requests. 

The fourth and fifth GPU pipeline execution models in VersaPipe are "Coarse pipeline" and "Hybrid pipeline". In "Coarse pipeline", each pipeline stage is bounded to one SM. In "Hybrid pipeline", each pipeline stage is assigned to multiple thread blocks on a few SMs. As discussed before, a synthesized FPGA is not able to perform different functions based on the SM id. Therefore these two execution models are not feasible to FPGA. 

\section{MKPipe: A Compiler Framework for Multi-Kernel Workloads}
\label{sec:MKPIPE}
\subsection{Overview}
\begin{figure}[htbp]
  \centering
  \includegraphics[height=8.0cm,width=7.5cm]{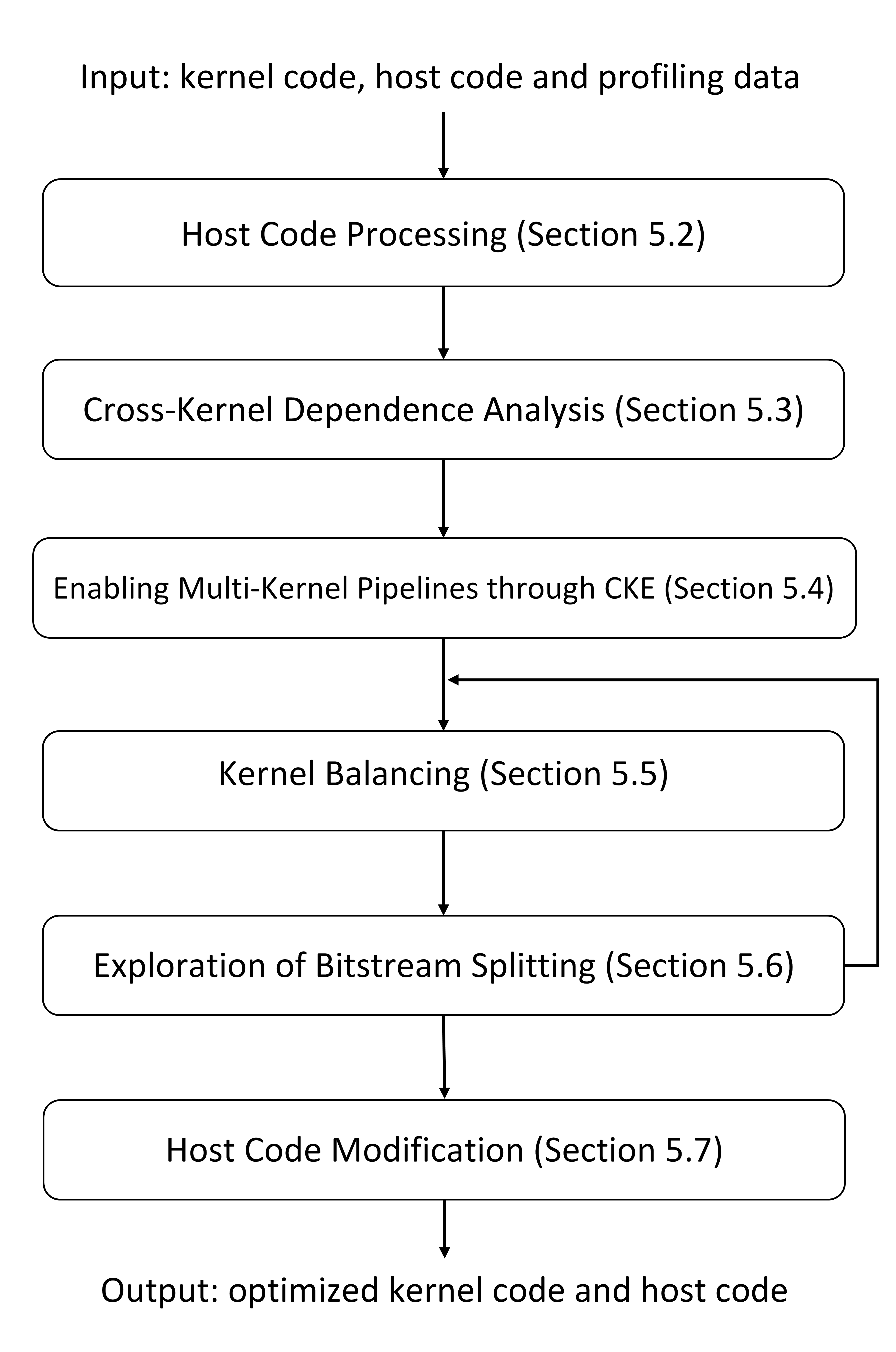}
  \caption{MKPipe: our proposed compiler framework}
  \label{framework}
\end{figure}

Our compiler framework is shown in Figure~\ref{framework}. 
The input to our compiler includes naive kernel code, host code, and profiling data. The naive kernel code means that there is no device-specific or platform-specific optimization. 
In our implementation, the naive kernel is the same as the one from the benchmark suite with all the optimization \#pragma and attributes stripped. The profiling data include the execution time and throughput of each naive kernel, where the throughput of one kernel is computed as the ratio of the output data size over the execution time. The compiler generates the kernel data flow graph from the host code and determines how the kernels can be executed concurrently while satisfying data dependency. Then the compiler analyzes the producer-consumer relationship between workitems/loop iterations in different kernels (workitems for NDRange kernels or loop iterations for single-workitem kernels) and uses different ways to enable CKE. Next, kernel balancing is performed to either balance the throughput in a multi-kernel pipeline or adjust the resource allocation among the kernels which require global synchronization. After kernel balancing, the compiler explores the option of bitstream splitting. Finally, the compiler produces the optimized kernel code and the host code.

\subsection{Host Code Processing}
\label{sec:hostcodeprocessing}
The compiler derives the kernel data flow graph from the host code. The kernels are invoked in the host code using {\em clEnqueueTask} or {\em clEnqueueNDRangeKernel} functions. Their inputs and outputs arguments are explicitly set in {\em clSetKernelArg} functions. Among the kernels, the compiler excludes the kernels that can not be executed concurrently using the condition that they have dependency carried over through CPUs or CPU memory. 

\subsection{Cross-Kernel Dependency Analysis}
\label{sec:cross-kerneldependency}
For kernels with data dependency, the compiler analyzes the kernel code to identify the producer-consumer relationship among their workitems/loop iterations for NDRange/single-workitem kernels, respectively. As the data dependency is carried over through the variables with the same name, for each global-memory variable, the compiler searches all the kernels to see which one(s) uses it as live-in/live-out. As the array indices in OpenCL workloads are typically affine functions of workitem ids or loop iteration indices, the compiler performs polyhedra analysis~\cite{dragonBook} to determine the exact dependency between the workitems/iterations of the producer kernel and those of the consumer kernel. Based on this producer-consumer relationship, the dependency between two kernels at the workitem/iteration level is classified into the following categories: few-to-few, few-to-many, many-to-many and many-to-few.

For example, in the code of the two single-workitem naive kernels of the CFD benchmark as shown in Figure~\ref{fig:cfdnaive}, the compiler finds that the global variable 'fluxes\_energy' is produced in the kernel 'compute\_flux' and consumed in the kernel 'time\_step'. When the iteration index variable i == j, these two iterations in the two kernels access the same global memory address. Therefore, the compiler identifies this producer-consumer relationship as one-to-one (or few-to-few), since one loop iteration in the producer kernel produces the data for one iteration in the consumer kernel. 
\begin{figure}[htbp]
\begin{lstlisting}[columns=flexible,frame=tlrb,basicstyle=\small,mathescape]
__kernel compute_flux(..., __global float* fluxes_energy){
    for (int i = 0; i < nelr; ++i) {
      fluxes_energy[i] = flux_i_density_energy;}}
__kernel time_step(..., __global float* fluxes_energy){
    for (int j = 0; j < nelr; ++j) {
      v_energy[j] = old_v_energy[j] + factor $\times$ fluxes_energy[j];}}

\end{lstlisting}
\caption{Code segment of the CFD benchmark}
\label{fig:cfdnaive}
\end{figure}

Another example is shown in Figure~\ref{fig:ludnaive}, where the two NDRange kernels in the LUD benchmark access a common array 'm'. In each workitem of the kernel 'lud\_perimeter', a number (BSIZE) of elements in array 'm' are updated as shown in line 6 of Figure~\ref{fig:ludnaive} and their array indices are a linear function of its workitem id and its workgroup id. In the kernel 'lud\_internal', each workitem reads two elements from the same array for computation and their indices are also linear functions of their workitem ids and workgroup ids. Through polyhedral analysis, the compiler determines the dependency relationship between the producer workitem/workgroup ids in the kernel 'lud\_perimeter' and the consumer workitem/workgroup ids in the kernel 'lud\_internal' and classifies the producer-consumer relationship as one-to-many (or few-to-many). 

Besides the dependency relationship, the compiler also produces a constant queue structure, id\_queue, which is used to determine the desired execution order for the workitems in the consumer kernel. This id\_queue is used in the workitem/workgroup id remapping step (Section~\ref{sec:id remapping}). As the workitems in the producer kernel are dispatched in the sequential order based on their workitem ids, the compiler mimics this order to process the workitems of the producer kernel. For each producer workitem, the compiler checks its dependent workitems in the consumer kernel. If a dependent workitem has its dependency completely resolved, its workitem id will be pushed into the id\_queue. If there are multiple dependent workitems in the consumer kernel and they are ready at the same time, all their workitem ids will be pushed in the id\_queue. The compiler also builds a similar queue at the workgroup granularity, i.e., the queue contains the consumer workgroup ids. 
\subsection{Enabling Multi-Kernel Pipelining}
\label{sec:buidlingCKE}
\begin{figure}[htbp]
  \centering
  \includegraphics[height=6.0cm,width=8cm]{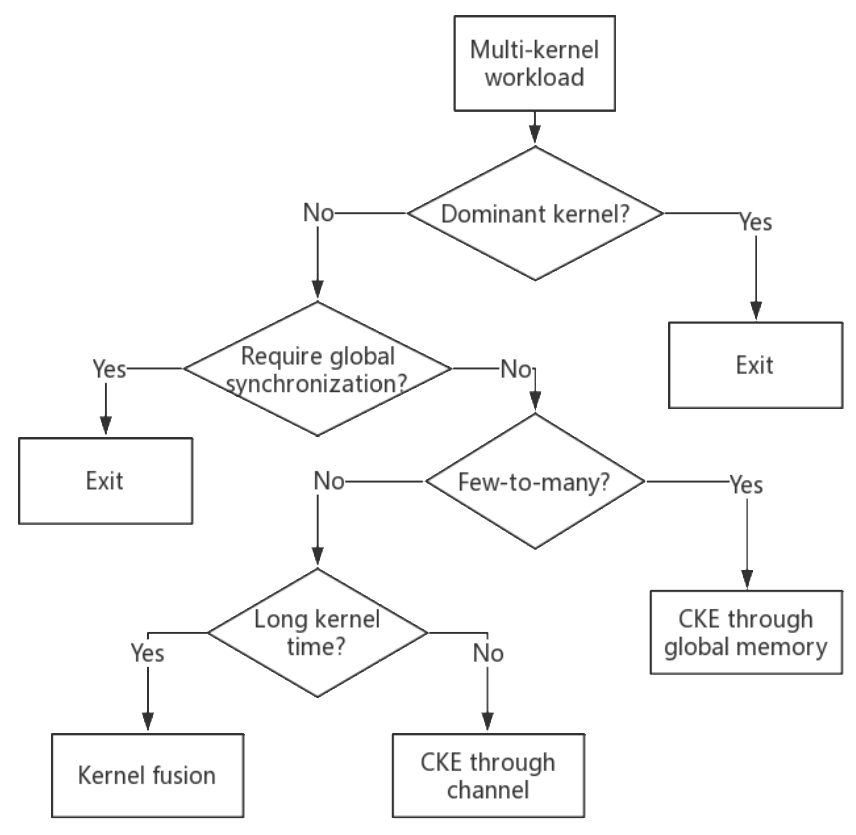}
  \caption{A systematic approach for enabling multi-kernel pipelines.}
      \vspace*{-4mm}
  \label{systematicapproach}
\end{figure}

Based on the producer-consumer relationship between the kernels, we propose a systematic decision tree approach to enable multi-kernel pipelines through CKE. 
Our approach is shown in Figure~\ref{systematicapproach}. First, the compiler checks if there is a dominant kernel in the workload. We define a kernel as dominant if its execution time is over 95\% of total execution time. The reason for such a check is that as long as this dominant kernel has high resource utilization, the overall utilization is high and CKE would have very limited impact. Then, the compiler checks if there is a need for global synchronization among the kernels as a result of the producer-consumer relationship. For many-to-many or many-to-few producer-consumer relationship, the consumer workitems/loop iterations have to wait for almost all the producer workitems/iterations to finish. Therefore, the gains from CKE typically is not high enough to offset the potential overhead of CKE. As a result, global synchronizations are justified in such cases. 

For multi-kernels not requiring global synchronization, the compiler explores different ways to enable multi-kernel pipelines through CKE. If the kernels exhibit few-to-many producer-consumer relationship, we propose to use global memory for data communication and this approach is referred to as CKE through global memory. If they exhibit few-to-few producer-consumer relationship, the compiler chooses between kernel fusion and CKE through channels. It estimates the overall execution time. When the execution time is high, the compiler chooses kernel fusion and CKE with channel otherwise for the reason discussed in Section~\ref{sec:CKEwithChannel}.

\subsubsection{Kernel Fusion}
\label{sec:fusion}
Kernel fusion fuses multiple kernels into a single one. It can lead to a longer pipeline and exploit better pipeline-level parallelism across kernels. For kernels in the single-workitem mode, fusion can be done by simply merging the kernel code without change. For kernels in the NDRange mode, fusion is also straightforward for the compiler as long as they share the same workgroup size and the same number of workgroups. If not, fusion becomes challenging for the compiler.
Therefore, our compiler would not fuse such NDRange kernels (i.e., kernels with different workgroup sizes) and resorts to CKE with channel instead.

As an example, as seen from the naive single-workitem kernel code of the benchmark CFD in Figure~\ref{fig:cfdnaive}, the two kernels have to communicate data through global memory, which incurs high performance overhead. Kernel fusion eliminates this problem as seen in Figure~\ref{fig:cfdfusion}, which contains the code after the compiler merges the kernels. After the two loops are fused by the compiler through the classical loop fusion optimization, the redundant global memory accesses to the array 'fluxes\_energy' is eliminated as this array is not a final live out of the application. 
\begin{figure}[htbp]
\begin{lstlisting}[columns=fullflexible,frame=tlrb,basicstyle=\small,mathescape]
__kernel compute_flux_time_step(...){
  for (int i = 0; i < nelr; ++i){
    v_energy[i] = old_v_energy[i] + factor$\times$flux_i_density_energy;
}}

\end{lstlisting}
    \vspace*{-2mm}
\caption{Code segment of CFD after the compiler applies kernel fusion}
\label{fig:cfdfusion}
    \vspace*{-2mm}
\end{figure}

From this example, we can see that besides enabling pipelining, kernel fusion can reduce resource consumption. 

On the other hand, we found some limitations of kernel fusion, which are also justifications for using multi-kernel designs over a single monolithic kernel. The first is the rigid requirement on the identical number of workitems/iterations in the producer and consumer kernel as discussed above. The second is that a single large kernel loses the benefit of design modularity and code reuse. The third is that a single kernel loses the flexibility to optimize different kernels differently. For example, with multiple kernels, the compiler can apply compute unit replication only for a particular kernel. Such selective optimization would become quite difficult once the kernels are fused. 

\subsubsection{CKE with Channel}
\label{sec:CKEwithChannel}
Similar to kernel fusion, using channels could also remove global memory reads/writes. Based on the producer-consumer relationship, the compiler introduces the code for defining channels and replaces global memory reads/writes with channel reads/writes. In the CFD example, Figure~\ref{fig:cfdcode} shows the code after the compiler performs the CKE with channel optimization.
\begin{figure}[htbp]
\begin{lstlisting}[columns=fullflexible,frame=tlrb,basicstyle=\small,mathescape]
channel float c_energy;
__kernel compute_flux(...){
  for (int i = 0; i < nelr; ++i){
    flux_i_energy = flux_i_density_energy;
    write_channel(c_energy,flux_i_energy);
}}
__kernel time_step(...){
  for (int j = 0; j < nelr; ++j){
    v_energy[j] = old_v_energy[j]+factor$\times$read_channel(c_energy);
}}

\end{lstlisting}
    \vspace*{-2mm}
\caption{Code segment of CFD after the compiler applies CKE with channels}
    \vspace*{-2mm}
\label{fig:cfdcode}
\end{figure}

CKE with channel is more flexible than kernel fusion as it is not limited by the strict requirement on the same number of workitems/iteration in the producer and consumer kernel. 

\begin{figure}[htbp]
  \centering
  \includegraphics[height=2.2cm,width=8.5cm]{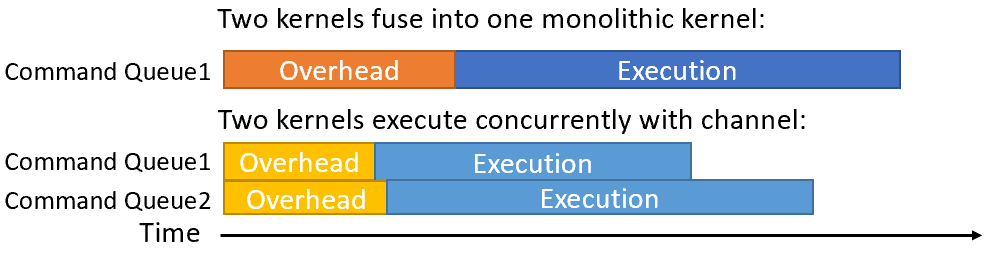}
  \caption{Difference in kernel invocation overhead of kernel fusion and CKE with channel}
      \vspace*{-2mm}
  \label{fig:overhead}
\end{figure}

Our results show that the compiler generated hardware designs from either the channel version or the fused kernel version are quite similar. The only difference is that some on-chip wiring or pipeline registers are replaced with FIFOs. 

Another distinguishing benefit of the CKE through channel over kernel fusion is the opportunity to reduce kernel launching overhead. Figure~\ref{fig:overhead} illustrates this with an example. With multiple kernels and each kernel in a different command queue, the kernel invocations overlap with each other. In comparison, the fused kernel has higher launching overhead due to its aggregated resource usage and a greater number of kernel arguments. This kernel invocation overhead trade off has not been studied in previous works. This reduction is more evident when the overall execution time is short and less evident otherwise. Therefore, the compiler favors CKE through channels for kernels with low execution time, as shown in Figure~\ref{systematicapproach}.

\subsubsection{CKE with Global Memory}

For kernels with their workitems/iterations having few-to-many producer-consumer relationship, we propose to enable CKE with global memory. For NDRange kernels in this category, the compiler introduces an array as global flags for the workitems in the producer kernel. This flag array is initialized to 0. When a producer workitem finishes its assigned work, it sets the corresponding global flag (i.e., array element indexed with the workitem id) to 1. The compiler inserts code in the consumer kernel such that the workitems in consumer kernel will wait until the corresponding flag is set to 1, indicating the data has been updated by the workitems in the producer kernel. For single workitem kernels, the same procedure applies except that the iterations replace the workitems. 

We use the LUD benchmark to illustrate CKE with global memory. Figure~\ref{fig:ludnaive} and~\ref{fig:ludglobal} shows the code before and after the compiler performs the optimization. First, the compiler inserts a global array 'flag'. In the producer kernel, each workitem sets the flag using its workitem id as the index (line 10 in Figure~\ref{fig:ludglobal}) after the updates to the array 'm'. A fence is added by the compiler in line 9 in Figure~\ref{fig:ludglobal} to ensure the correct memory update order. In the consumer kernel, the compiler introduces the flag check for each read site of the array 'm' and generates the code for accessing the flag of the producer workitem based on the workitem dependency relationship determined during the dependency analysis step. Such code is shown in lines 15, 16, 19 and 22 in Figure~\ref{fig:ludglobal}.
\subsubsection{Workitem/Workgroup ID Remapping}
\label{sec:id remapping}
 The execution order of the work-items (or iterations) in NDRange (or single-workitem) kernels depends on the hardware and may not match our desired order. Our empirical results show that for each kernel, work-items with increasing ids (or iterations with increasing iterator value) are dispatched in the sequential order. If there's only one compute unit, the work-groups with increasing workgroup ids will also be executed in the sequential order. However, such a rigid order may not match the dependence resolution order between the producers and consumers.

\begin{figure}[htbp]
    \begin{subfigure}{}
    \begin{lstlisting}[columns=fullflexible,frame=tlrb,basicstyle=\small,mathescape]
__kernel lud_perimeter(__global float* m,int mat_dim,int offset){
  int tx = get_local_id(0), bx = get_group_id(0);
  peri_row_array_offset = offset + (bx + 1)$\times$ BSIZE
  for (int i = 0; i < BSIZE; ++i){
    ...
    m[peri_row_array_offset + tx] = peri_row[tx $\times$ BSIZE + i];
    peri_row_aray_offset += mat_dim; }}
__kernel lud_internal(__global float* m,int mat_dim,int offset){
  int bx = get_group_id(0), by = get_group_id(1);
  int tx = get_local_id(0), ty = get_local_id(1);
  int global_row_id = (by + 1) $\times$ BSIZE;
  int global_col_id = (bx + 1) $\times$ BSIZE;
  peri_row[ty $\times$ BSIZE + tx] = m[offset + ty $\times$ mat_dim + 
  global_col_id + tx];
  peri_col[ty $\times$ BSIZE + tx] = m[offset + (ty + global_row_id) 
  $\times$ mat_dim + tx]; ...}}
    \end{lstlisting}
           \vspace*{-4mm}
   \caption{Code segment of the Naive LUD benchmark}
    \label{fig:ludnaive}
    \end{subfigure}
    \begin{subfigure}{}
    \begin{lstlisting}[columns=fullflexible,frame=tlrb,basicstyle=\small,mathescape]
__kernel lud_perimeter(__global float* m,int mat_dim,int offset){
  int tx = get_local_id(0);
  int bx = get_group_id(0);
  peri_row_array_offset = offset + (bx + 1)$\times$ BSIZE
  for (int i = 0; i < BSIZE; ++i){
    ...
    m[peri_row_array_offset + tx] = peri_row[tx $\times$ BSIZE + i];
    peri_row_aray_offset += mat_dim;}
  mem_fence(CLK_GLOBAL_MEM_FENCE);
  flag[bx $\times$ group_size_1 + tx] = 1; }
__kernel lud_internal(__global float* m, int mat_dim, int offset,
  __global int* flag){
  int bx = get_group_id(0), by = get_group_id(1);
  int tx = get_local_id(0), ty = get_local_id(1);
  int wait_id_1 = bx $\times$ group_size_1 + tx;
  int wait_id_2 = by $\times$ group_size_1 + tx;
  int global_row_id = (by + 1) $\times$ BSIZE;
  int global_col_id = (bx + 1) $\times$ BSIZE;
  while(!flag[wait_id_1]){}
  peri_row[ty $\times$ BSIZE + tx] = m[offset + ty $\times$ mat_dim +
  global_col_id + tx];
  while(!flag[wait_id_2]){}
  peri_col[ty $\times$ BSIZE + tx] = m[offset + (ty + global_row_id) 
  $\times$ mat_dim + tx]; ...}}

    \end{lstlisting}
           \vspace*{-4mm}
    \caption{Code segment of LUD after CKE with global memory}
    \label{fig:ludglobal}
    \end{subfigure}
\end{figure}
 
\begin{figure}[htbp]
  \centering
  \includegraphics[height=3.5cm,width=8cm]{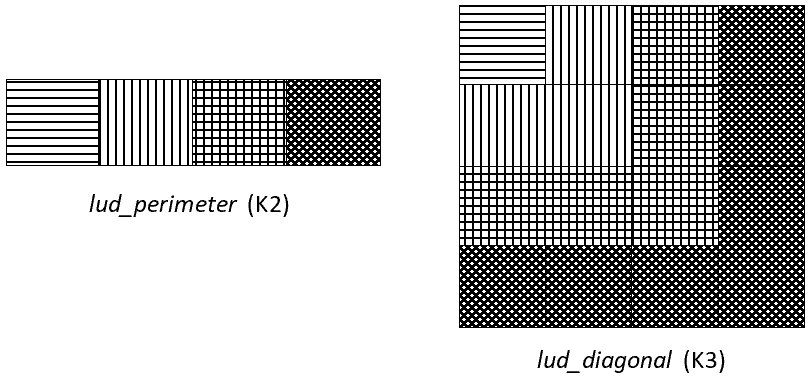}
  \caption{Data dependency between the workgroups in K2 and K3 of the LUD benchmark.}
      \vspace*{-4mm}
  \label{fig:luddependency}
\end{figure}

Figure~\ref{fig:luddependency} shows the data dependency between the workgroups of the producer kernel and the workgroups in the consumer kernel in the LUD benchmark. One block in the figure represents one work-group. The blocks with the same pattern have data dependence. From the figure, we can see that the workgroup 0 in the kernel 'lud\_perimeter' produces the data for the workgroup (0,0) in the kernel 'lud\_diagonal'; the workgroup 1 in the kernel 'lud\_perimeter' produces the data for the workgroups (0,1), (1,0), (1,1) in the kernel 'lud\_diagonal', etc. As the default execution order of the workgroups in the consumer kernel 'lud\_diagonal' is workgroup (0,0), (0,1), (0,2), (0,3), etc., the workgroups (0,2) and (0,3) will have to wait for their data to be produced although the workgroups (1,0) and (1,1) already have their data ready. 

In order to resolve this execution order mismatch problem, we propose a work-item/workgroup id remapping approach. The observation is that we can reassign each work-item/work-group a new id to change the execution order. Here, our compiler makes use of the id\_queue structures produced during the dependency analysis step. This queue structure is stored in the constant memory to take advantage of the FPGA on-chip constant cache. Each consumer workgroup/workitem reads the queue using their workgroup/workitem id as the index. To explore different options, our compiler produces three versions of code, no id remapping, workgroup id remapping only, workgroup id remapping and workitem id remapping. These three versions are synthesized and tested to select the best performing one. 
\begin{figure}[htbp]
\begin{lstlisting}[columns=fullflexible,frame=tlrb,basicstyle=\small,mathescape]
__kernel lud_internal(__global float* m, int mat_dim, int offset, 
__global int* flag, int* id_queue_bx,  int* id_queue_by){
  int bx = get_group_id(0), by = get_group_id(1);
  int tx = get_local_id(0), ty = get_local_id(1);
  int wait_id_1 = bx $\times$ group_size_1 + tx;
  int wait_id_2 = by $\times$ group_size_1 + tx;
  bx = id_queue_bx[bx], by = id_queue_by[by];
  int global_row_id = (by + 1) $\times$ BSIZE;
  int global_col_id = (bx + 1) $\times$ BSIZE;
  while(!flag[wait_id_1]){}
  peri_row[ty $\times$ BSIZE + tx] = m[offset + ty $\times$ mat_dim + 
  global_col_id + tx];
  while(!flag[wait_id_2]){}
  peri_col[ty $\times$ BSIZE + tx] = m[offset + (ty + global_row_id) 
  $\times$ mat_dim + tx]; ...}}

\end{lstlisting}
    \vspace*{-4mm}
\caption{Code segment of the LUD benchmark after workgroup id remapping}
    \vspace*{-4mm}
\label{fig:reordering}
\end{figure}

Figure~\ref{fig:reordering} shows the code of the consumer kernel 'lud\_internal' of LUD after the compiler applies the workgroup id remapping. The constant workgroup\_id queues are in the kernel parameters and the id remapping code is shown in line 7 in Figure~\ref{fig:reordering}. 

\subsection{Kernel Balancing} 
\label{sec:kernelbalancing}
With multiple kernels sharing the FPGA device, we need to coordinate the optimizations upon them. Our compiler considers two different scenarios for kernel balancing. The first is for kernels with CKE, i.e., the kernels form a pipeline. The second is that there are global synchronizations separating kernels and making them run sequentially.  

For kernels with CKE, each kernel becomes a stage in a multi-kernel pipeline. Therefore, the goal is to balance the throughput among the stages. However, if kernels are separated by global synchronizations, throughput imbalance is not an issue. The goal then is to balance resource allocation such that more resources can be allocated for the kernels, which can lower the most execution time from such additional resources. A workload may also contain both cases at the same time. For example, the CFD benchmark has three kernels. Our compiler determines that it is beneficial to enable CKE between K2 and K3 while K1 should be ended with a global synchronization. In such a case, the compiler considers the K2\&3 pipeline as a single kernel and allocates resources to K1 and K2\&3 accordingly using the algorithm discussed in Section~\ref{sec:resourcebalancing}. Then, the allocated resources for K2\&3 are further distributed between K2 and K3 using the algorithm discussed in Section~\ref{sec:throughput balancing} for throughput balancing.

In our compiler framework, we consider three parameters for single-kernel optimizations: compute unit replication (CU) factor, SIMD factor, and loop unroll (Unroll) factor. As these factors have similar performance impacts, i.e., increasing any factor by N times can potentially increase the throughput by N times, our compiler first determines a unified performance factor, denoted as \textit{N$_{uni}$}, for each kernel, and then realize this factor by adjusting the unroll, SIMD, and CU factors as discussed in Section~\ref{sec:optimization parameters}.

\subsubsection{Throughput Balancing}
\label{sec:throughput balancing}
When kernels are running concurrently in a pipeline, the throughput of the pipeline is limited by the stage with the lowest throughput. Therefore, we propose an approach to assign resources gradually to different kernels. The algorithm is shown in Algorithm \ref{alg:throughputsteps}. The algorithm takes the throughputs of the naive kernels, \textit{Tp$_{1...k}$}, as input, which are obtained during the profiling step. The algorithm iteratively searches for the kernel with the lowest throughput and increases its unified performance factor by 1 each time. Then, the optimization parameters, i.e., Unroll factors, SIMD factors, and CU factors, are derived and the kernel code is generated. Next, we resort to the OpenCL compiler to estimate the static resource consumption based on the updated kernel code. For the dynamic bandwidth resource, we assume the utilization is the bandwidth of the naive kernel times the unified performance factor. The process repeats until one of the resources becomes fully utilized. Note that in this algorithm, the OpenCL compiler is not used to fully synthesize the hardware. Instead, it is only used to generate the resource estimate, which can be quickly finished. As we do not synthesize the actual hardware, the throughput of a kernel with a unified performance factor, \textit{N$_{uni}$}, is estimated as \textit{N$_{uni}$} times the throughput of the naive kernel, i.e., \textit{N$_{uni}$xT$_{p}$}, as shown in line 3 of Algorithm \ref{alg:throughputsteps}. 

Since the throughputs and the resource utilization for different performance factors are estimated, we add an auto-tuning step to compensate for potential estimation errors after the algorithm determines the performance factors of each kernel. During auto-tuning, based on \textit{N$_{uni}$} computed for each kernel, we compile \& synthesize multiple designs for an limited range of performance factors [\textit{N$_{uni}$} $\pm$ \textit{p}] to search the best \textit{N$_{uni\_opt}$}. The search space is determined through a user-defined parameter \textit{p}.  

\begin{algorithm}[h]
\DontPrintSemicolon
  \KwInput{\textit{T$_{p1...k}$}:naive throughput for each kernel;}
  \KwOutput{\textit{N$_{uni1...k}$}:unified factor for each kernel;}
  \KwData{\textit{TP$_{1...k}$}:calculated throughput for each kernel;}
  $\textit{N$_{uni1...k}$}\gets 1$\; 
   \While{total resource estimation $\leq 100\%$}
   {
   		$TP_{1...k} \gets N_{uni1...k}\times T_{p1...k};$\;
   		\textit{Find kernel j with lowest throughput TP;}\;
   	   	$N_{unij} \gets N_{unij} + 1;$ //$\times 2$ if  SIMD  is  used\;   
   	   	\textit{Calculate corresponding}  $N_{unroll}, N_{SIMD}$ \textit{and} $N_{CU}$ \textit{with algorithms in Figure~\ref{fig:tuning};}\;
   	   	\textit{Generate the code for kernel j;}\;
   	   	\textit{Extract resource estimation of kernel j from the log file generated by the OpenCL compiler;}\;
   }
\caption{Algorithm of computing \textit{N$_{uni}$} for throughput balancing among kernels in a pipeline.}
\label{alg:throughputsteps}
\end{algorithm}

\subsubsection{Resource Balancing}
\label{sec:resourcebalancing}
For the kernels that are separated with global synchronizations, we distribute the resources to kernels according to their performance impact. We propose an iterative approach, as shown in Algorithm \ref{alg:resourcesteps}. The algorithm takes the execution time of the naive kernels \textit{T$_{1...k}$} as input and determines the unified performance factor for each kernel. In each step, The compiler computes the performance impact of resource allocation for each kernel as the ratio of performance improvement over the change in the critical resource utilization when the unified performance factor is increased by 1. The changes in static resource allocation for each kernel $\Delta U_{1...k}$ is obtained from the log file generated from the OpenCL compiler. The updated dynamic bandwidth utilization is assumed as (the bandwidth of the naive kernel x \textit{N$_{uni}$}). The performance improvement for each kernel, when its unified performance factor, \textit{N$_{uni}$}, is increased by 1, is estimated as: $\frac{T_{1...k}}{N_{uni1...k}}-\frac{T_{1...k}}{N_{uni1...k} + 1}$, which is equal to $\frac{T_{1...k}}{N_{uni1...k}(N_{uni1...k}+1)}$, as used in line 4 of Algorithm \ref{alg:resourcesteps}. Then, the kernel with the highest performance impact from additional resource will have its unified performance factor incremented, i.e., the resources granted. This process repeats until the critical resource is fully utilized. 

After the performance factors are determined from the algorithm, an auto-tuning process, similar to it discussed in Section~\ref{sec:throughput balancing}, is used to fine-tune these factors in a user-defined range. 

\subsubsection{Determining Optimization Parameters}
\label{sec:optimization parameters}
After a kernel is assigned a unified performance factor, \textit{N$_{uni}$}, the single-kernel optimization parameters, the Unroll factor, the SIMD factor, and the CU factor are adjusted to realize it. Among these factors, loop unrolling has the lowest resource consumption, and compute unit replication has the highest. Therefore, our compiler determines these three factors following this order. The pseudo-code of the algorithm is shown in Figure~\ref{fig:tuning}. In the figure, the constant MAX\_UNROLL\_ FACTOR is the maximum iterations in a loop. The boolean constant VEC represents whether the kernel code is beneficial from vectorization/SIMD and such information is obtained during the profiling phase of each kernel. As shown in Figure~\ref{fig:tuning}, the Unroll factor is tried first to realize the unified performance factor \textit{N$_{uni}$}. If it cannot, the SIMD factor is considered and the last is the CU factor. Since the OpenCL for FPGA compiler requires that SIMD factors should be a power of 2, if SIMD factor is chosen, the unified performance factor, \textit{N$_{uni}$}, should be doubled rather than being increased by one, as shown in the comment of line 6 in Algorithm 1 and line 7 in Algorithm 2.
\begin{algorithm}[h]
\DontPrintSemicolon
 \KwInput{\textit{T$_{1...k}$}:naive execution for each kernel;}
  \KwOutput{\textit{N$_{uni1...k}$}:unified factor for each kernel;}
  \KwData{\textit{U$_{1...k}$}:resource utilization for each kernel;}
  $\textit{N$_{uni1...k}$}\gets 1$\; 
   \While{total resource estimation $\leq 100\%$}
   {
   		\textit{Check critical resource type and derive }$\Delta U_{1...k}$\textit{ for that type of resource from resource estimation}\;
   		$\Delta T_{1...k} \gets \frac{T_{1...k}}{N_{uni1...k}(N_{uni1...k}+1)};$\;
   	   	\textit{Find kernel j with highest }$\frac{\Delta T_{j}}{\Delta U_{j}}$\; 
   	   	$N_{unij} \gets N_{unij} + 1; $//$\times 2$  if  SIMD  is  used \;
   	   	\textit{Calculate corresponding}  $N_{unroll}, N_{SIMD}$ \textit{and} $N_{CU}$ \textit{with algorithms in Figure~\ref{fig:tuning};}\;
   	   	\textit{Derive total resource estimation from log file;}\;
   }
\caption{Algorithm of computing \textit{N$_{uni}$} for resource balancing among multiple kernels separated with global synchronizations.}
\label{alg:resourcesteps}
\end{algorithm}

\begin{figure}[htbp]
\begin{lstlisting}[columns=fullflexible,basicstyle=\small, frame=tlrb,mathescape]
if (N$_{uni}$ < MAX_UNROLL_FACTOR) {
    N$_{unroll}$ = N$_{uni}$;}
else if (N$_{uni}$ $\geq$ MAX_UNROLL_FACTOR && VEC) {
    N$_{SIMD}$ = N$_{uni}$ / MAX_UNROLL_FACTOR;
    N$_{unroll}$ = MAX_UNROLL_FACTOR;}
else {
    N$_{CU}$ = N$_{uni}$ / MAX_UNROLL_FACTOR;
    N$_{unroll}$ = MAX_UNROLL_FACTOR;}

\end{lstlisting}
\caption{Pseudo-code for determining optimization parameters from a unified performance factor, \textit{N$_{uni}$}.}
\label{fig:tuning}
\end{figure}

\subsection{Bitstream Splitting}
The bitstream splitting optimization explores the option of placing kernels in multiple bitstream files. This way, more resources are available for each kernel such that more aggressive single-kernel optimizations can be performed. However, using multiple bitstreams has to pay the penalty of device reprogramming and data transfer between the device and the host. Therefore, we limit the maximal number of bitstreams as 2. As a result, if there are more than two kernels in a workload, our compiler decouples them into two virtual kernels. Such decoupling is essentially the same as bi-partitioning the kernel data flow graph. 

Our compiler employs the following criteria for bi-partitioning the graph: (a) loops cannot be partitioned unless each iteration of the loop has very high execution latency compared to reprogramming overhead; (b) a multi-kernel pipeline can not be broken by partitioning; and (c) the difference between the accumulated critical resource utilization over time in either partition needs to be minimized. A loop in the kernel data flow graph means that the kernels will be invoked multiple times. If we break kernels in a loop into different bitstreams, we have to pay the reprogramming overhead for each iteration. Therefore, unless the execution time of each iteration is high, the loop should not be partitioned. The last criterion aims to isolate the long-running kernels, which are resource constrained due to co-residence with other kernels. Such kernels are more likely to benefit from more resources. Using the notation of Equation 2, the condition can be expressed as find a partition to minimize $|\textit{T$_{1}$} \times\textit{ERU$_{1}$} - \textit{T$_{2}$}\times\textit{ERU$_{2}$}|$.
As the number of the kernels in multi-kernel pipelines are small, our compiler exhaustively goes through all the possible partitions to find one that meets the criteria. 

With the two virtual kernels, our compiler uses Equation 2 to determine whether to put them into separate bitstreams or to let them co-reside in the same one. In the equation, \textit{K$_{1}$} and \textit{K$_{2}$} are the two virtual kernels. The ERU of them are \textit{ERU$_{1}$} and \textit{ERU$_{2}$} and their execution times are \textit{T$_{1}$} and \textit{T$_{2}$}. The reprogramming and data transfer overhead are \textit{T$_{r}$} and \textit{T$_{d}$}, respectively. We consider kernel co-residence in a single bitstream beneficial if:
\begin{equation}
\label{eq1}
\textit{T$_{1}$} + \textit{T$_{2}$}<\textit{T$_{1}$} \times \textit{ERU$_{1}$} + \textit{T$_{2}$}\times\textit{ERU$_{2}$} + \textit{T$_{r}$} + \textit{T$_{d}$}
\end{equation}
The LHS of Equation 2 is the execution time if both kernels reside on the same device. The RHS is an estimate of the execution time if they are separated into two different bitstreams. When one kernel monopolizes the device, its execution time can be reduced with more aggressive optimizations. Such reduced execution time is estimated with a factor of the kernel's ERU, i.e, the utilization of its critical resource. For example, if one kernel uses 80\% of the DSP blocks, when the entire chip, i.e., 100\% DSP blocks are available to it, the potential performance improvement would be 100\%/80\%. The corresponding execution time is $80\% \times $T, i.e., $ERU\times$T. 
If LHS is less than RHS, co-residence is preferred. Otherwise, the compiler produces two source code files, one for each virtual kernel, which will be used by the OpenCL compiler to synthesize into separate bitstreams.

We tested the reprogramming overhead \textit{T$_{r}$} using kernels with different complexities. We found that the reprogramming overhead is around 1400ms for different kernels and it is independent upon the complexity or resource requirement of the kernel. 

\subsection{Host Code Modification}
After kernel optimizations, the host code is adjusted accordingly. For kernel fusion, unnecessary kernel invocations and allocations for the global memory data that are used for cross kernel communication would be removed. For CKE with channel and CKE with global memory, kernel arguments are adjusted. The compiler also allocates global memory for the global 'flags' array and the 'id\_order' array.
All the \textit{clFinish} functions between concurrent executing kernels are removed since they are synchronization points. 

\subsection{Compilation Overhead}
The main cost of MKPipe is the OpenCL to FPGA compilation overhead. We will compile the kernels during two steps: the profiling step and auto-tuning step. Since the naive kernel has no optimization pragma and attributes, the compilation time in profiling step is usually much smaller than compiling the baseline kernels as they enable these pragma/attributes. During the auto-tuning step, each kernel needs to be compiled for $2p + 1$ times, where $p$ is the user-defined parameter discussed in Section~\ref{sec:throughput balancing}. Nonetheless, all these compilations in the tuning step can be performed in parallel.

\section{Methodology}
We implement our proposed compiler framework as a source-to-source compiler. Our compiler takes advantage of Clang, the front end of LLVM~\cite{Lattner:Mthesis}. Specifically, it leverages the ASTMatcher and ASTTransformer in Clang for source code analysis and transformations. We used the Candl tool~\cite{bastoul2008candl} for the polyhedral analysis. The user-defined parameter, \textit{p} discussed in Section~\ref{sec:kernelbalancing}, is set to 2.
We studied the multi-kernel workloads (a total of 6) that are already optimized for FPGA in Spector \cite{gautier2016spector}, Rodinia \cite{zohouri2016evaluating} and OpenDwarf\cite{verma2016opendwarf} benchmark suites and two multi-kernel workloads from an irregular graph benchmark for GPU, Pannotia\cite{che2013pannotia}. Table \ref{table:benchmarkcharact} summarizes the key characteristics of the benchmarks. 

Our experiments are performed with Altera OpenCL SDK18.1 which is the latest version supported by Terasic's DE5-Net board. The board has 4GB DDR3 memory and a Stratix V GX FPGA. 
\begin{scriptsize}
\begin{table}[h!]
  \centering
  \small
  \begin{adjustbox}{width=0.48\textwidth}
  \begin{tabular}{|c|c|c|c|}
    \hline
    \textbf{Benchmark}& \textbf{Key Characteristics} & \textbf{ Key Optimization} \\
    \hline
    \hline
    BFS\cite{gautier2016spector} & Dominant kernel & Kernel balancing \\
    \hline
    Hist\cite{gautier2016spector} & One-to-one & Kernel fusion  \\
    \hline
    CFD\cite{zohouri2016evaluating} & One-to-one &CKE with channels \\
    \hline
    LUD\cite{zohouri2016evaluating} & One-to-many &CKE with global memory\\
    \hline
    BP\cite{zohouri2016evaluating} & Splitting beneficial& Bitstream splitting \\
    \hline
    Tdm\cite{verma2016opendwarf} & Dependency through CPU & Kernel balancing\\
    \hline
    Coloring\cite{che2013pannotia} & One-to-one & kernel fusion\\
    \hline
    Dijkstra\cite{che2013pannotia} & One-to-one & CKE with channels\\
    \hline
  \end{tabular}
  \end{adjustbox}
  \caption{Benchmarks used in our experiments.}
          \vspace*{-8mm}
  \label{table:benchmarkcharact}
\end{table}
\end{scriptsize}
\section{Evaluation}
\subsection{Overall Results}
\begin{figure}[htbp]
  \centering
  \includegraphics[height=4.5cm,width=8.5cm]{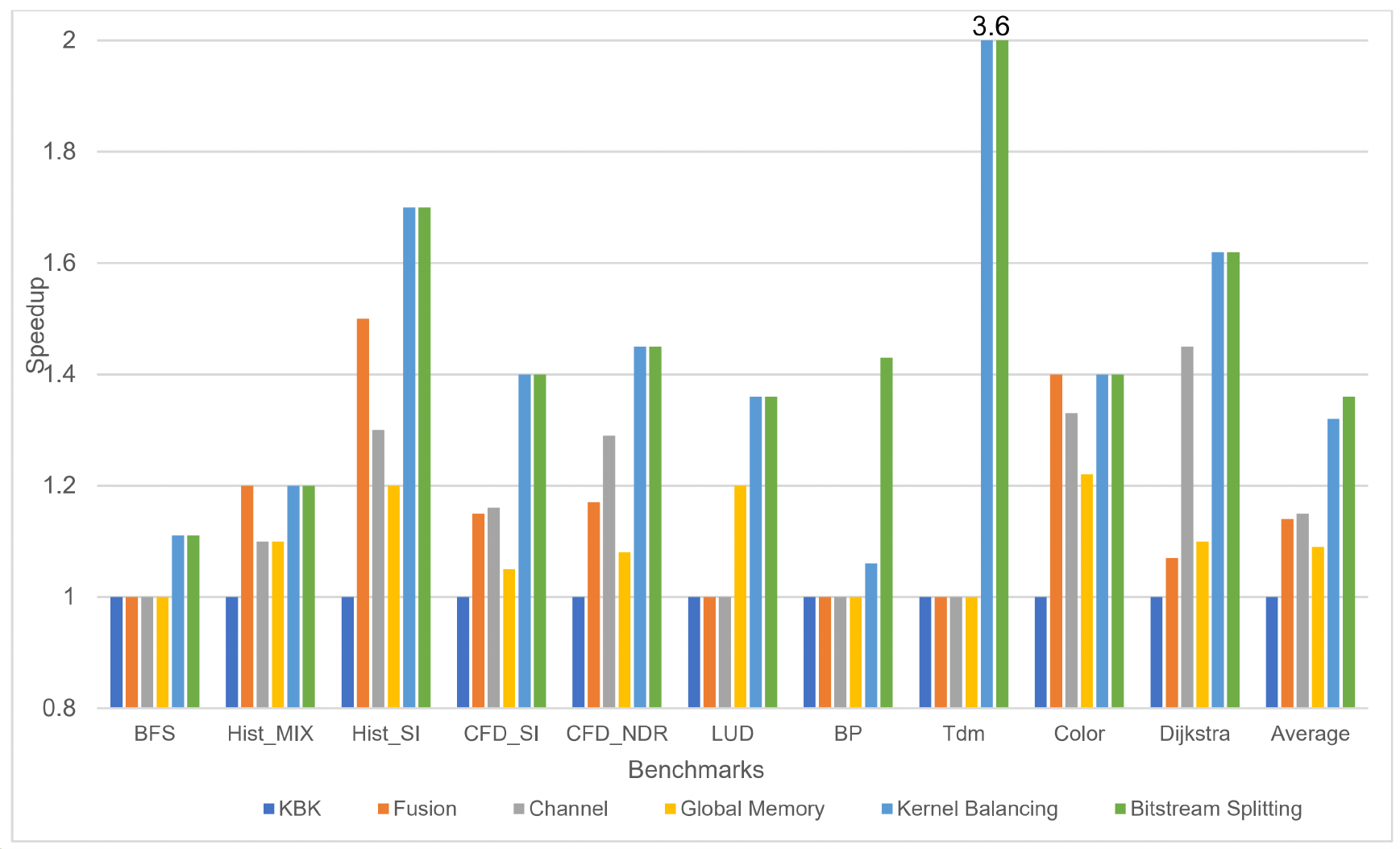}
  \caption{Impact of kernel execution model and optimization steps}
        \vspace*{-4mm}
  \label{fig:speedup}
\end{figure}

Figure~\ref{fig:speedup} reports the normalized performance of the multi-kernel workloads. We use the following notations: 'KBK' represents the kernels from the benchmark suites which use the KBK model. KBK is our baseline. For the benchmarks derived from the GPU benchmark suite, we applied the SIMD and CU attributes to optimize the kernel as our baseline. 'Fusion' represents kernels executed with a hybrid model of 'KBK' and 'RTC'. The reason is that the kernel fusion model or RTC does not support global synchronization between stages. As a result, the kernels with global synchronization in between are executed with the KBK model. 'Channel' represents kernels executed with the hybrid model of 'KBK' and 'CKE with channel'. 
'Global Memory' represents kernels executed with the hybrid model of 'KBK' and 'CKE with global memory'. 'Kernel Balancing' shows the speedup of the kernels with the best pipeline model and kernel balancing optimization. 'Bitstream splitting' is for the kernels optimized with kernel balancing and bitstream splitting. 

Among all the benchmarks, BP uses single workitem kernels. CFD has kernel implementations in both single-workitem (labeled 'CFD \_SI') and NDRange mode (labeled 'CFD\_NDR'), and Hist has the original implementation using an NDRange producer kernel and a single workitem consumer kernel (labeled 'Hist\_MIX'). We found that this NDRange kernel results in low frequency and rewrote it as a single workitem kernel. This Hist version is labeled 'Hist\_SI'. All the remaining benchmarks use NDRange kernels.

\begin{scriptsize}
\begin{table*}[h!]
  \centering
\begin{adjustbox}{width=1\textwidth}
  \begin{tabular}{|c|c|c|c|c|c|c|c|c|c|c|c|c|c|c|c|c|c|c|c|c|c|c|c|}
    \hline
    \small
    & \multicolumn{2}{c|}{BFS} & \multicolumn{2}{c|}{Hist\_MIX}&
    \multicolumn{2}{c|}{Hist\_SI}&
    \multicolumn{2}{c|}{CFD\_SI}&
    \multicolumn{2}{c|}{CFD\_NDR}&
    \multicolumn{2}{c|}{LUD}
    & \multicolumn{3}{c|}{BP}
    & \multicolumn{2}{c|}{Tdm}
     & \multicolumn{2}{c|}{Color}
     & \multicolumn{2}{c|}{Dijkstra}\\
    \hline
    \textbf{Resource}
  & Base & Opt & Base & Opt  & Base & Opt & Base & Opt & Base & Opt  & Base & Opt & Base & Opt1 & Opt2  & Base & Opt &Base & Opt & Base & Opt\\
    \hline
    \hline
    ALUTs (\%)
    & 27 & 33 & 15 & 15 & 18 & 15 & 49 & 46 & 45 & 71 & 60 & 61 & 25& 32& 30& 23 & 45 & 53 & 62 & 44 & 52\\
   \hline
    FFs (\%)
    & 21 & 26 & 11 & 12 & 15 & 11 & 25 & 23 & 24 & 35 & 25 & 45 & 22 & 24 & 30 & 16 & 36 & 25 & 33 & 28 & 31\\
   \hline
    RAMs (\%)
    & 54 & 68 & 85 & 87 & 57 & 25 & 54 & 48 & 50 & 62 & 72 & 83& 40 & 35 & 44 & 36 & 85 & 68 & 76 & 59 & 64\\
   \hline
    DSPs (\%)
    & 0 & 0 & 1 & 1 & 1 & 0 & 63 & 63 & 63 & 91 & 74 & 80 & 31 & 77 & 56 & 2 & 6 & 0 & 0 &0 & 0\\
    \hline
    \hline
    Frequency (MHz)
    & 217 & 211 & 202 & 194 & 220 & 230 & 225 & 228 & 226 & 225 & 229 & 227 & 228 & 213 & 226 & 221 & 208 & 265 & 225 & 260 & 232\\

   \hline
  \end{tabular}
  \end{adjustbox}
  \caption{Resource consumption of all benchmarks. Opt1,2 are the two bitstreams resulted from bitstream splitting.}
        \vspace*{-4mm}
  \label{table:all resource}
\end{table*}
\end{scriptsize}

From Figure~\ref{fig:speedup}, we can see that the CKE optimization and the kernel balancing optimization contribute the most performance improvement. Overall, the multi-kernel workloads optimized by MKPipe achieve up to 3.6x (1.4x on average) speedup over the baseline.
Among the multi-kernel workloads, BFS has a dominant kernel, which takes 95.8\% of the overall execution time. MKPipe identifies this dominant kernel and performs kernel balancing optimization. Our optimized kernel achieves a speedup of 1.1x as our compiler balances the optimizations on the kernels more judiciously. The Histogram benchmark has one producer kernel and one consumer kernel, their dependency relationship is identified as one-to-one. For the single workitem implementation (Hist\_SI), MKPipe generates both fused design and CKE with channel design. As the fused design forms a longer loop body, the OpenCL compiler optimizes the code more effectively and the synthesized design achieves a speedup of 1.7x over the baseline. For HIST\_MIX, due to different numbers of workitems in the producer and consumer kernels, MKPipe chooses to enable CKE using channels. The benchmark Tdm benefits the most from the kernel balancing optimization as it efficiently searches a large design space of the optimization parameters. The main benefit of LUD comes from CKE with global memory and workgroup mapping as discussed in Section~\ref{sec:MKPIPE}. 
Color benefits from kernel fusion. Dijkstra benefits from CKE with channel due to the low execution time of its kernels. For the remaining benchmarks, CFD and BP, we analyze them in Section 7.3.

Besides the performance impact, we present the resource consumption and the frequency of different designs for each benchmark in Table \ref{table:all resource}. From the table, we can see that for most benchmarks our optimized design utilizes resources more aggressively and one side effect is the slightly lower frequency due to longer critical paths.    
\subsection{Comparison with GPU}
\begin{figure}[htbp]
  \centering
  \includegraphics[height=3.5cm,width=8.5cm]{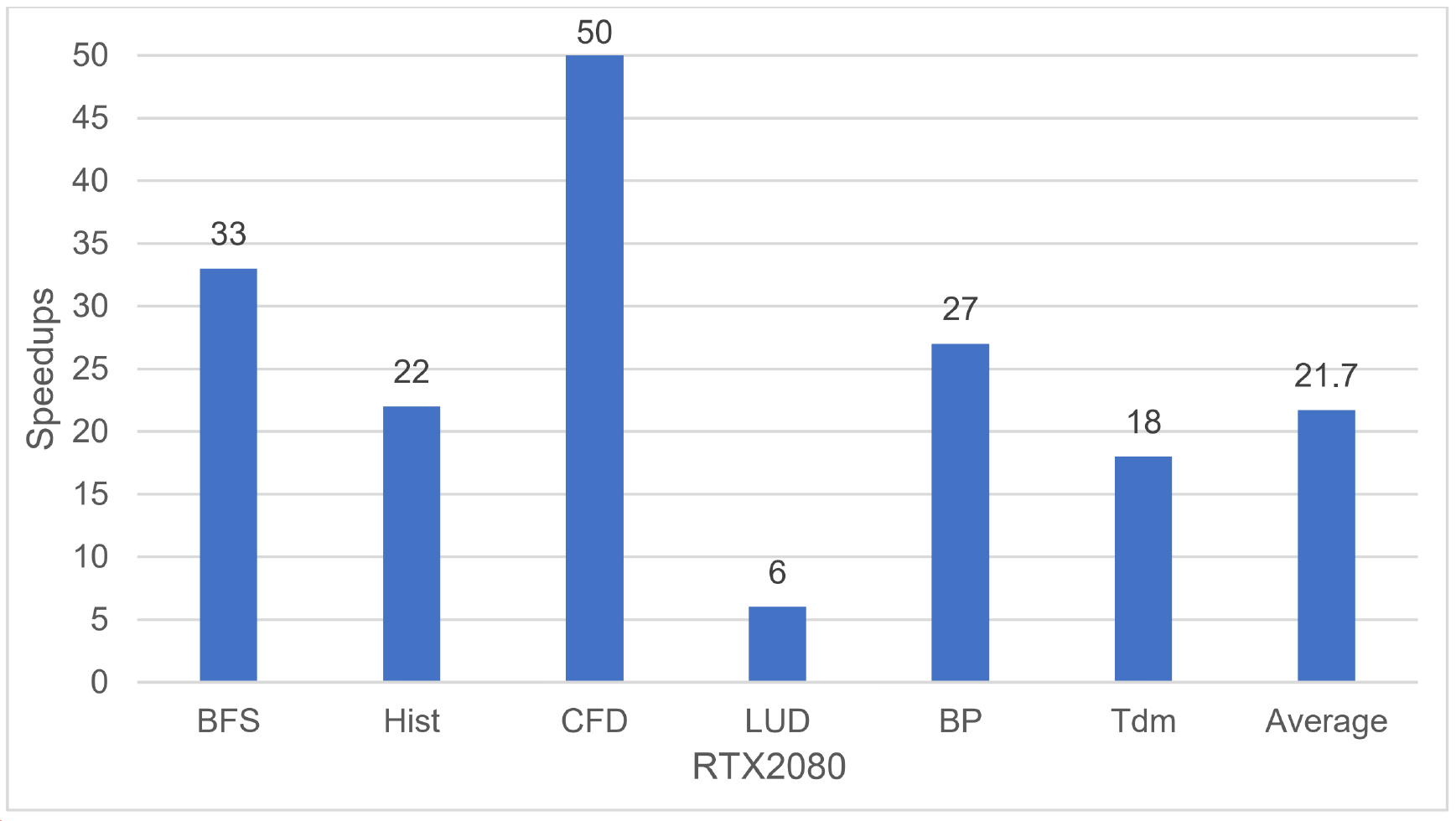}
      \vspace*{-6mm}
  \caption{Speedup of RTX2080 compared to Stratix V}
    \vspace*{-6mm}
  \label{fig:GPU}
\end{figure}

In this experiment, we compare the FPGA performance with NVIDIA RTX 2080 GPUs. For benchmarks BFS and Hist, the CUDA kernels from Parboil~\cite{stratton2012parboil} benchmark suite are used as they are optimized for GPUs. Similarly, for benchmarks CFD, LUD and BP, the CUDA kernels from Rodinia benchmark suite are used. For benchmark Tdm, the OpenCL kernels is used as the OpenDwarf benchmark suite does not have the CUDA version. The benchmarks from Pannotia benchmark suite are not included since they require AMD drivers and SDK support.
The results are shown in Figure~\ref{fig:GPU}. 
Given the bandwidth difference between our FPGA board (25.6 GB/s for Stratix V) and GPU (448 GB/s on RTX2080), the performance of OpenCL kernels for FPGA is not competitive. 

To make a more fair comparison, we include a performance projection for the state-of-art Stratix 10 MX FPGA. Compared to Stratix V GX, Stratix 10 MX~\cite{stratix10} has 6x DSP capability, 2.6x memory blocks 20x memory bandwidth(512GB/s). Taking the advantage of 14nm manufacturing node and HyperFlex technology~\cite{hutton2015hyperflex}, Stratix 10 family FPGA boards are expected to reach an operating frequency $f_{max}$ of up to 1 GHz. However, when $f_{max}$ is limited by the critical path, HyperFlex will have a limited impact. Therefore, we only assume a 150MHz increase in $f_{max}$ compared to Stratix V, which is in accordance with prior study~\cite{zohouri2018combined}~\cite{chung2018servingdnn}. As can be seen in Table~\ref{table:all resource}, most of the optimized benchmarks are bandwidth limited. 
Based on the existing performance estimation model~\cite{wang2016performance}, the speedup of benchmarks on Stratix 10 MX can be predicted as:
\begin{equation}
\label{effectiveresource}
\small
Speedup = \frac{freq_{proj}}{freq} \times \frac{\#Banks_{proj}}{\#Banks} \times \frac{\#Mem\_trans\_width_{proj}}{\#Mem\_trans\_width}
\end{equation}
$freq_{proj}$ and $freq$ are the frequencies of Stratix 10 MX and Stratix V. $\#Banks_{proj}$ and $\#Banks$ are the number of memory banks which are 32 and 2 for Stratix 10 MX and Stratix V, respectively. \textit{\#Mem \_trans \_width} is the maximum transaction width and it is 64Byptes for both devices. Based on these data, the average speedup (geometric mean) of all six benchmarks is 26.8x. As can be found in figure~\ref{fig:GPU}, the average speedup of kernels on Stratix 10 MX FPGA is comparable with the average speedup of the kernels on the state-of-the-art GPU. Such results are also consistent with existing works~\cite{zohouri2016evaluating,zohouri2018combined} that FPGAs deliver inferior performance but superior energy efficiency to the same generation GPUs.
      \vspace*{-4mm}

\subsection{Case Studies}
\subsubsection{CFD}
The kernel data flow graph of CFD is shown in Figure~\ref{fig:cfd_invocation}. Since K2 and K3 form an inner loop, MKPipe chooses to enable concurrent execution between K2 and K3. After cross-kernel dependency analysis, MKPipe identifies the producer-consumer relationship between K2 and K3 as one-to-one as discussed in Section~\ref{sec:cross-kerneldependency}.

\begin{figure}[htbp]
  \centering
  \includegraphics[height=3cm,width=8cm]{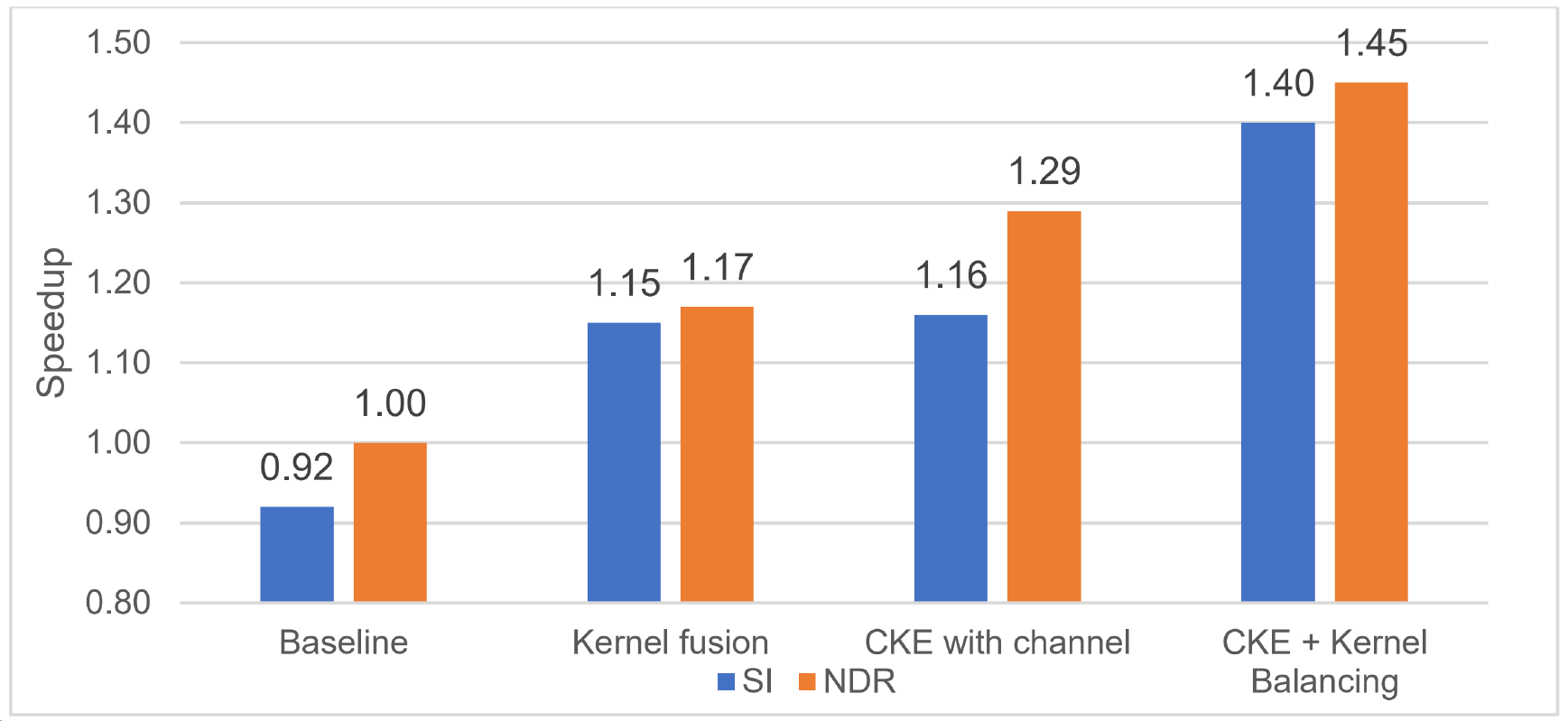}
  \caption{Speedups of optimized kernels over baseline. SI: single work-item kernels; NDR: NDRange kernels}
      \vspace*{-4mm}
  \label{fig:cfdspeedup}
\end{figure}

Since CFD has two versions, one using single-workitem kernels and the other using NDRange ones. We show their performance after each optimization step in Figure~\ref{fig:cfdspeedup}. Between fusion and CKE with channel, MKPipe picks CKE with channel due to the short execution time. After optimizations, especially kernel balancing, the optimized NDRange implementation achieves the highest performance. 

\subsubsection{BP}
\begin{figure}[htbp]
  \centering
  \includegraphics[height=1.3cm, width=8cm]{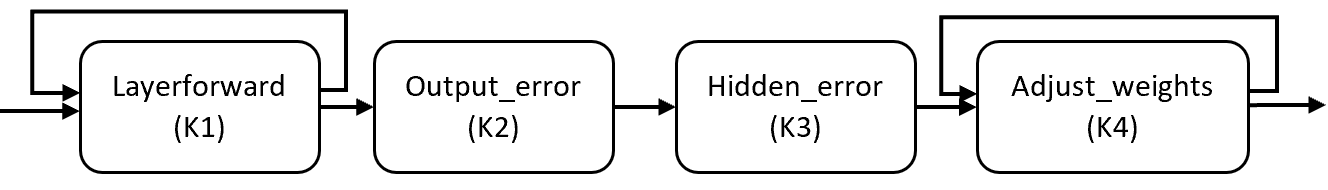}
  \caption{Kernel data flow graph of BP.}
      \vspace*{-4mm}
  \label{fig:bp_invocation}
\end{figure}
The backpropagation (BP) benchmark trains the weights in a layered neural network. It has four kernels and 
the kernel data flow graph is showed in Figure~\ref{fig:bp_invocation}. The profiling data show the first kernel invocations and the last kernel invocations take 20\% and 76\% of the overall execution time, respectively. Given the loops in the kernel data flow graph, MKPipe mainly applies the resource balancing and kernel splitting optimizations. During the bitstream splitting step, MKPipe partitions K4 from the rest kernels due to its long execution time and its relatively high ERU. After the kernels are put in separate bitstreams, the kernel balancing step is repeated such that both kernel K1 and K4 are more aggressively optimized. The reduced execution time from K1 and K4 over-weighs the reprogramming overhead and a significant net gain (1.43x) in performance is achieved.

\section{Conclusions}
In this paper, we present a source-to-source compiler framework, MKPipe, for optimizing multi-kernel workloads in OpenCL for FPGA. There are two key optimizations. One is to enable multi-kernel pipelining through different ways of concurrent kernel execution (CKE). The other is to adaptively balance the throughput or the resource among the multiple kernels. The key novelty of this work is: (a) a systematic compiler optimization scheme to enable multi-kernel pipelines; (b) CKE through global memory along with workitem/workgroup id remapping; (c) algorithms to balance the throughput and/or resource consumption among the kernels in a multi-kernel pipeline; and (d) a new approach to explore the option of bitstream splitting.


\bibliographystyle{ACM-Reference-Format}
\balance
\bibliography{ref}

\appendix

\end{document}